\documentstyle[pra,floats,aps,twocolumn,epsfig]{revtex}
\begin{document}\bibliographystyle{apsrev}

\wideabs{
\title{Observation of Nonclassical Photon Statistics due to Quantum Interference}
\author{Y. J. Lu and Z. Y. Ou}
\address{Department of Physics, Indiana
University-Purdue University Indianapolis \\ 402 N Blackford
Street, Indianapolis, In 46202}
\date{June 28, 2001}
\maketitle

\widetext
\begin{abstract}
The nonclassical effect of photon anti-bunching is observed in the mixed field of
a narrow band two-photon source and a coherent field under certain condition. A variety of
different features in photon statistics are found to be the consequence of a two-photon
interference effect with dependence on the relative phase of the fields. Besides the
anti-bunching effect, we find another one of the features to be also nonclassical.  These
features emphasize the importance of quantum entanglement.
\end{abstract}

}

\narrowtext
Nonclassical photon statistics such as photon anti-bunching is usually observed in
two-level atomic system with resonant excitation, where quantum nature of the process
prevents the emission of two photons at the same time. Therefore, an anti-bunched photon field
will have less two-photon events than fields with random photon statistics such as a
coherent field from a laser.  Historically, this was the first observed nonclassical effect
requiring a full quantum description of light \cite{kimble}. Since then, such a nonclasical effect has
been observed in the fluorescence in a variety of systems consisting of a small number of atoms, ions and
molecules\cite{diedrich,schubert,basche}. Recently, potential application in quantum cryptography has
renewed the interest in producing anti-bunched photon source
\cite{demartini,kitson,brunel,fleury,kurtsiefer,brouri,kim,michler}. When pumped by a short pulse,
the systems mentioned above can form a photon gun \cite{law} and become a good single-photon source
\cite{brouri2}.

So far most proposals and realized systems are based on some atomic transitions. A
completely different mechanism was proposed by Stoler \cite{stoler} for the production of anti-bunched
photon field as early as in 1974. Stoler's
proposal is based on parametric amplifier and was realized experimentally by Koashi et al\cite{koashi1}.
The pulsed system was used in that experiment following Stoler's suggestion that anti-bunching may occur
only in the "transient" regime of the parametric amplifier.  Recently, however, it was suggested
\cite{vyas} that anti-bunched cw field can be generated in the homodyning of an optical parametric
oscillator (OPO) and a strong coherent local oscillator. Coherent
homodyne of down-converted field was first studied by Grangier et al. \cite{grangier} to investigate
nonlocal effect of two-photon system and demonstrated by Kuzmich et al \cite {kuz}. Similar scheme was
also realized by Koashi et al\cite{koashi2} in the pulsed system. Naively, it is hard to understand how
anti-bunched light can be generated from a highly bunched two-photon source such as parametric
down-conversion, where photons are emitted in pairs.  It turns out that the underlining principle is
quantum interference.   Although high two-photon events occur from parametric down-conversion, there are
also low accidental two-photon events from a coherent field. With a strong coherent field, the accidental
two-photon events may have the same rate as that from parametric down-conversion.  If the two processes
are coherent, two-photon interference will lead to sinusoidal modulation of two-photon coincidence as a
function of the relative phase. At certain phase when destructive two-photon interference occurs,
two-photon events in the mixed field will be less than a coherent field, resulting in photon
anti-bunching while at complementary phase, constructive interference will lead to photon bunching.

In this paper, we report an experimental implementation of the above idea. We have observed
a number of quite different features in photon statistical distribution as the relative
phase between the down-converted field and the coherent field is changed. The observed
features include photon antibunching and bunching and a new feature that also has
nonclassical implication.

\begin{figure}[tbp]
\centerline{\epsfxsize=3.3in \epsffile{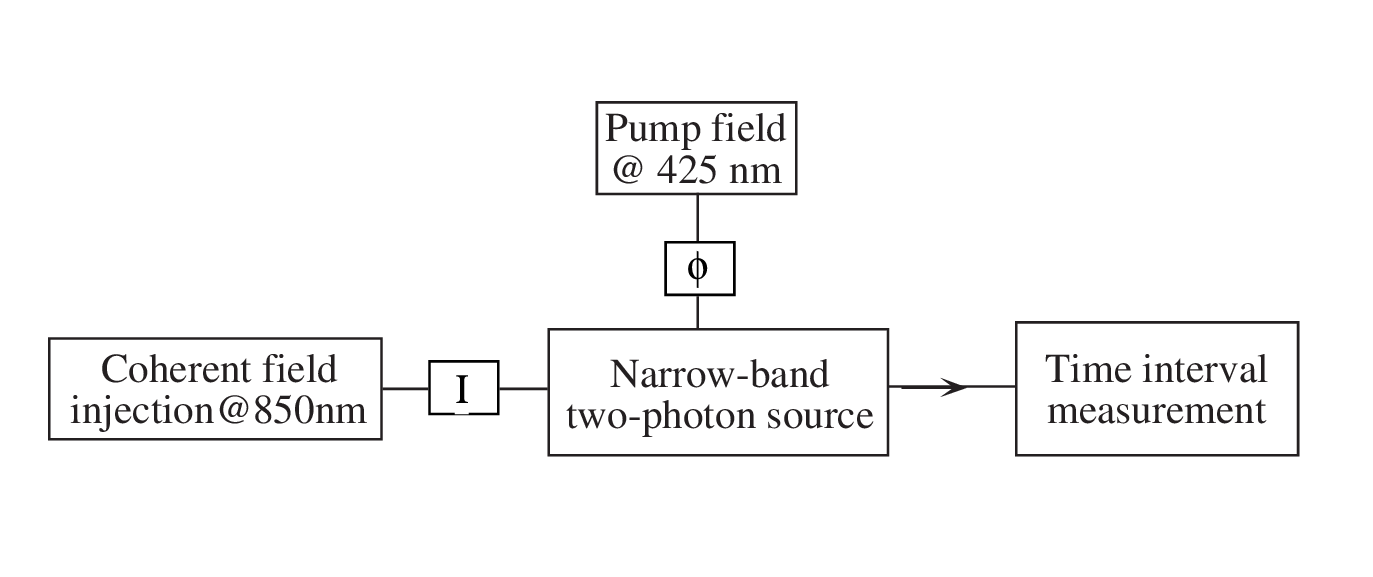}}
\caption{Sketch of the mixing of a coherent field and a two-photon
field.} \label{exp-setup}
\end{figure}

To understand more rigorously how this could happen, let us consider the mixing of a
 coherent state given by
\begin{eqnarray}
|\alpha\rangle  \approx |0\rangle + \alpha
|1\rangle + {\alpha^2\over\sqrt{2}} |2\rangle~~~~~(|\alpha| << 1)
\end{eqnarray}
with a
single-mode two-photon state from parametric down-conversion. For simplicity, we only treat them with a
simple single mode model. To be consistent with the experimental arrangement to be discussed later,
the mixing is done by injecting the coherent field into the parametric down-converter as shown in Fig.1.
Such a scheme is exactly same as the one proposed by Stoler \cite{stoler}. But we present it here again to
emphasize the role of quantum two-photon interference. We treat the coherent state as the initial state
for the down-converted mode (denoted by the creation operator
$\hat a$), which interacts with the pump field via Hamiltonian
\begin{eqnarray}
\hat H = j \hbar \eta V_p \hat a^{\dagger} \hat a^{\dagger} + h.c.,
\end{eqnarray}
where $V_p$ is the amplitude of the pump field and $\eta$ is some constant proportional to nonlinear
coefficient of the nonlinear medium.

The output state can be derived from the evolution operator
$\hat U (t) = exp
\{-j\hat H t/\hbar\}$ as
\begin{eqnarray}
|\Phi\rangle &=& \hat U(t) |\alpha\rangle\nonumber\\
&\approx &|0\rangle + \alpha |1\rangle + (\sqrt{2}\eta V_p t + \alpha^2/\sqrt{2})|2\rangle,
\end{eqnarray}
where $t$ is the interaction time that is proportional to the length of the nonlinear medium. In Eq.(3),
we made an expansion of the exponential function and dropped the higher-order terms by assuming $|\eta V_p
t| \sim |\alpha|^2 << 1$.  Therefore, the
single photon rate $P_1 \approx |\alpha|^2$ is same as that in Eq.(1) indicating that the mixed field is
dominated by the coherent field. The two-photon rate, however,  is given by
\begin{eqnarray}
P_2 = \big |\sqrt{2}\eta V_p t +
\alpha^2 /\sqrt{2}\big |^2,
\end{eqnarray}
which, with the selection of the relative phase between
the pump and the coherent fields, can be bigger than that of coherent field derived from Eq.(1) giving
rise to photon bunching effect or smaller for photon anti-bunching effect.  Especially when
$2\eta V_p t = - \alpha^2$, we have complete destructive two-photon interference with
zero rate for two-photon detection. Eq.(4) can be explained as follows: $\sqrt{2}\eta V_p t$ is
the two-photon amplitude for parametric down-conversion alone while $\alpha^2/\sqrt{2}$ is for the
coherent field; the addition of the two amplitudes gives the overall two-photon amplitude for the
combined field (the output field in Fig.1) resulting in Eq.(4) for two-photon rate.

Although the above simple single-mode picture makes it easier to understand the underlining principle of
the effect, in practice, the down-converted field has a wide spectrum while coherent field from a laser
has a narrow bandwidth. As demonstrated in our previous work in Ref.\cite{ou} and shown in the insets of
Fig.2, the wide-band down-converted field produces a bell-shaped temporal correlation for two-photon
coincidence measurement (solid line in the inset). For the coherent field, because of the randomness in
the arrival of photons, the temporal correlation function is simply a flat line (dash line in the inset).
The two functions provide the two-photon  rate of each field at some specific time delay, respectively.

When we mix the two fields, since the field is mostly dominated by the coherent field, there is no
significant single photon interference at one detector level. However, since two-photon rates for the two
fields are comparable at least at zero time delay, two-photon interference occurs
and is shown up in two-photon coincidence measurement with two detectors (intensity correlation). If the
amplitude of the coherent field is denoted  by $A$ and the
two-photon amplitude from down-converted field  by $F(\tau) \equiv B f(\tau)$ with
$B$ as a constant proportional to
the amplitude of the pump field and the interaction strength, and
\begin{eqnarray}
f(\tau) = {\Delta\omega_{C2} \Delta\omega_{opo} \over \Delta\omega_{opo} -\Delta\omega_{C2}}\bigg
(\frac{e^{-\tau
\Delta\omega_{C2}/2}}{\Delta\omega_{C2}} - \frac{e^{-\tau
\Delta\omega_{opo}/2}}{\Delta\omega_{opo}} \bigg )
\end{eqnarray}
obtained from Eq.(32) of Ref.\cite{ou}, the two-photon correlation function for the mixed field is
given by
\begin{eqnarray}
\Gamma^{(2)}(\tau)  &= &\big | A^2 + F(\tau) \big |^2,\nonumber \\
&=& |A|^4 + |B|^2 f^2(\tau) + 2 |A|^2 |Bf(\tau)| \cos \phi.
\end{eqnarray}
$\phi$ is the relative phase between the two fields. If we notice that
$f(\infty) = 0$, the normalized intensity correlation function is then (for stationary field only)
\begin{eqnarray}
g^{(2)}(\tau) \equiv {\Gamma^{(2)}(\tau) \over \Gamma^{(2)}(\infty) } = 1 + |b|^2 f^2(\tau) + 2 |bf(\tau)|
\cos \phi ,
\end{eqnarray}
where $b\equiv |B/A^2|$ gives the relative strength of the two fields. Eq.(7) presents the typical
phase dependent feature for interference effect and Fig.2 shows the correlation functions for various
parameters in Eqs.(5) and (7).

\begin{figure}[tbp]
\centerline{\epsfxsize=3.3in \epsffile{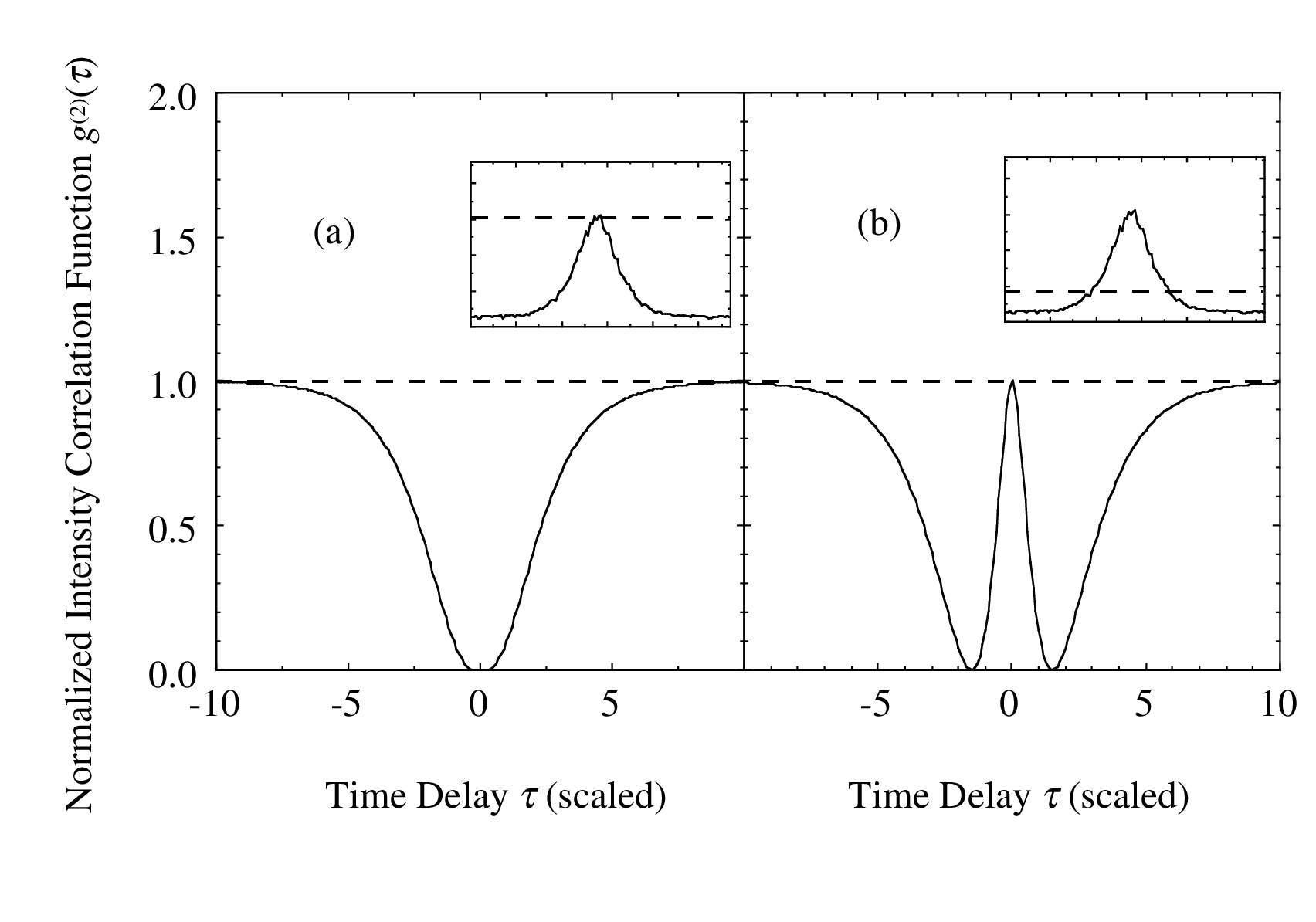}}
\caption{Normalized intensity (photon) correlation function
$g^{(2)}(\tau)$. Different types of photon anti-bunching effect
occur due to two-photon interference: complete cancellation occurs
(a) at zero time delay with $b = 1$ and $\phi = 180^{\circ}$ in
Eq.(7) and (b) at nonzero time delays with $b = 2$ and $\phi =
180^{\circ}$. $\tau$ is scaled by $4/ \Delta\omega_{opo}$ and
$\Delta\omega_{C2}/\Delta\omega_{opo} =2.8$ in Eq.(5).}
\label{Fig.2}
\end{figure}

The insets show the relative strength
in two-photon rate for the two fields, respectively.  Because two-photon rates are different
at different time delays, complete cancellation  occurs only at the locations where the two curves
intercept as shown in Fig.2. Fig.2(a) shows the usual type of anti-bunching effect; Fig.2(b), however,
shows a double dip feature with zero two-photon coincidence occurring at nonzero time delays. Similar
double dip shaped distribution can also be obtained at the same two-photon rate as in Fig.2(a) but with a
different relative phase.  All these
cases are nonclassical. In particular, Fig.2(a) violates the classical Schwartz inequalities
\begin{eqnarray}
g^{(2)}(0) \ge 1,~~~
g^{(2)}(0) \ge g^{(2)}(\tau),
\end{eqnarray}
and Fig.2(b) violates
\begin{eqnarray}
|g^{(2)}(0) - 1|\ge |g^{(2)}(\tau)-1|.
\end{eqnarray}
Although violations of inequalities in (8) were the most often observed nonclassical effects,
violation of inequality in (9) has only been observed recently\cite{foster}. In the following, we will
discuss the experimental procedures that lead to the observation of these nonclassical features in
normalized photon correlation function.

In the type of interference discussed above, both the coherent field and the two-photon field
are produced independent of each other. Therefore, the observation of the interference effect
requires that the detectors' response be faster than the fields' phase fluctuations. In other words,
since the phases of the two fields are uncorrelated, in order to observe the interference effect,
detection time must be short enough for stable phases. This requirement means that our detector's
response time must be shorter than the correlation time of the fields.  Recently, we
successfully built a narrow band two-photon source that satisfies this requirement\cite{ou}.

For the interference
experiment,  the layout is similar to that in Ref.\cite{ou} except that we
inject a coherent field split from the laser into the OPO cavity (C1 in Fig.1 of Ref.\cite{ou}) from the
left side (high reflector side) in consistency with the sketch in Fig.1 of current paper. The reason for
this arrangement is two-fold. Firstly, interference requires the presence of both fields at the same
time. But we use a mechanical chopper to eliminate the background from the auxiliary locking beam so that
the detected signal is not continuous (Even so, there still exists a rather large background  from
scattering). The current arrangement of letting the coherent field passing through the OPO cavity can
synchronize the two fields. Secondly, the OPO cavity can act as a spatial filter for the coherent field
to mode-match the down-converted field.

The time interval measurement for the two-photon event is accomplished by two avalanche
photo-detectors and a time-to-analog converter. The result is the intensity correlation function.
Typical results of the measurement are presented in Fig.3, which show the normalized intensity correlation
function at various relative phases and intensities. The adjustment of the relative phase is through the
change of the path of the pump field while for the relative intensity, it is done by adjusting the light
level of the injected coherent field. A clear evidence of photon anti-bunching is shown in Fig.3(a),
which is obtained at proper relative phase and relative intensity. Fig.3(b) is obtained with the relative
phase changed by $180^{\circ}$ from that of Fig.3(a). The double dip feature shows up in Fig.3(c) at a
different ratio of the two-photon rates of the two fields from Fig.3(a) but with same relative phase as
that of Fig.3(a).

It should be emphasized before the analysis of the experimental results that the photon
correlation function in Eq.(7) is extremely sensitive to the phases of the pump field and the coherent
field. A swing of the relative phase from
$180^{\circ}$ to $90^{\circ}$  (rather than to $0^{\circ}$) will cause the combined field to go from
anti-bunching to bunching. Unfortunately, the pump field and the injected coherent field travel through
different paths via multiple mirror sets so that their phase fluctuations are not correlated and to make
things worse, the mechanical chopper, which is used to reduce scattered light, stirs up an air flow around
the paths and disturbs the phases of the fields.  We do not have an active servo system to hold the phase
at the desired value.  So the phase may fluctuate and sometimes wander away from the desired value.
During the experiment, we manually hold the phase of the pump field at various values and our measurement
time is short. Typical period of measurement is about 5 sec, which is the reason behind the poor
statistics in Fig.3.  Even this cannot defeat the fast phase fluctuation but may prevent the slow drift
of the phase.  Because of the fast phase fluctuations, we cannot directly apply Eq.(7).  An average over
the phase difference $\phi$ must be performed to account for the phase fluctuations. The results are
shown as the solid curves in Fig.3.

\begin{figure}[tbp]
\centerline{\epsfxsize=3.3in \epsffile{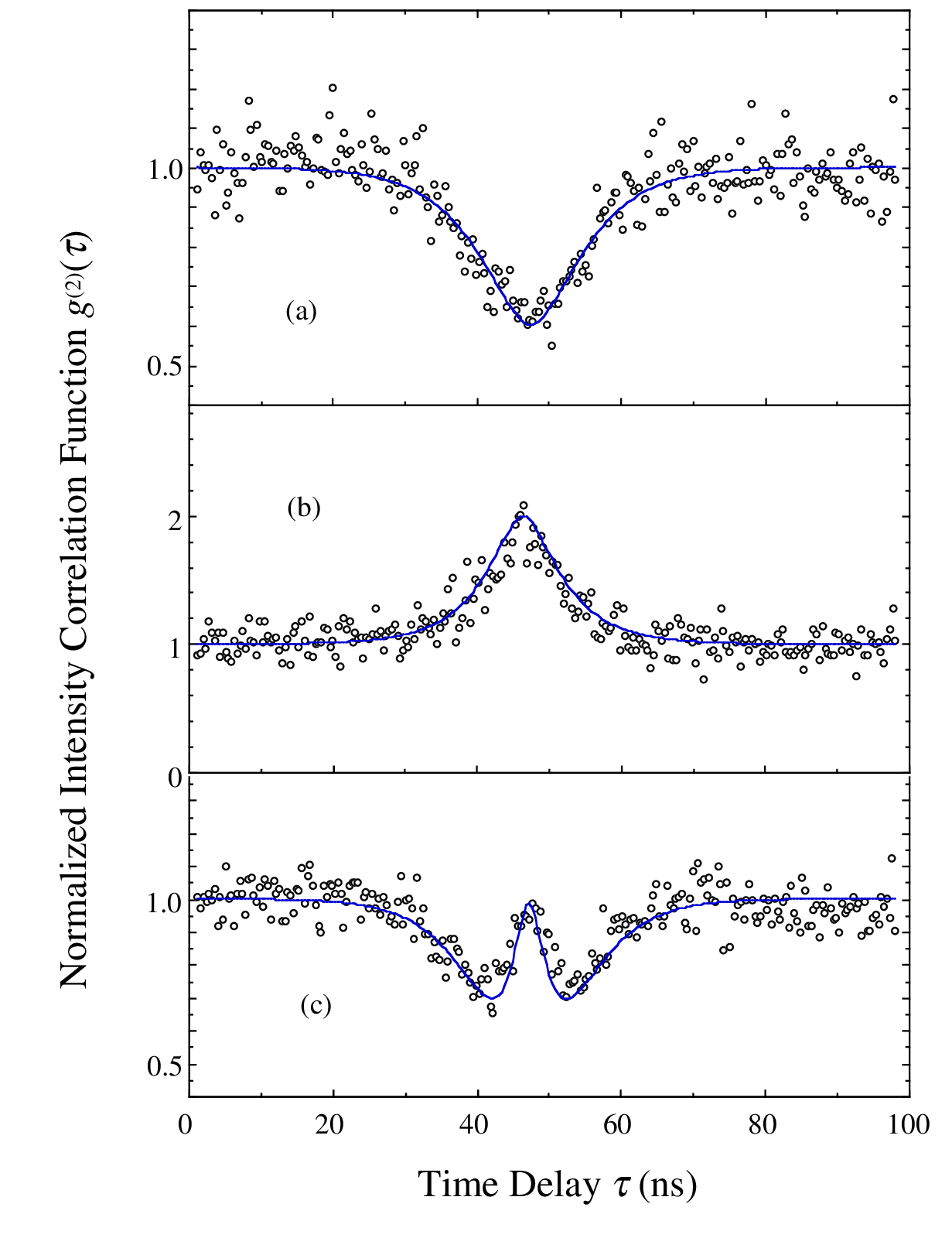}}
\caption{Measured normalized intensity correlation function
$g^{(2)}(\tau)$ at different conditions: (a) phase and relative
intensity are adjusted for maximum anti-bunching effect at zero
delay time ($\tau = 47.2 ns$); (b) phase is changed by
$180^{\circ}$; (c) phase is same as (a) but the intensity of the
coherent field is lowered.} \label{exp-setup}
\end{figure}

Obviously, Fig.3(a) shows the normal type of anti-bunching effect, where we have $g^{(2)}(0) < 1$ and
$g^{(2)}(0) < g^{(2)}(\tau)$. The suppression of two-photon coincidence at zero time delay is due to a
destructive interference. The distribution in Fig.3(c) violates the
classical inequality in (9). Fig.3(b), showing the typical photon bunching effect,
has an enhanced two-photon coincidence at zero delay time and is due to constructive interference. The
observed nonclassical effects are not as large as predicted in Fig.2. We believe the problem mostly lies
in the fluctuation of the phases of the fields involved. Another cause for the problem comes from
stimulated emission, which gives rise to the higher order terms that are omitted in Eq.(3). The higher
order terms do not participate in the interference process and thus provides a background that will raise
the level of the baseline in Fig.3 making the visibility of the interference small.  This problem can be
solved by lowering the pump level. But doing so will decrease the signal level and make the time longer
for data taking   so that the phase fluctuation becomes worse. A trade-off thus has to be made in our
experiment.  The decrease in visibility can also come from the background due to scattering from the
auxiliary locking beam.

An interesting feature in this experiment is that the intensity of
the two-photon source is much smaller than that of the coherent source by a factor of about 10. This is
another example of a weak nonclassical source making a strong effect on a strong classical source.
Although the scheme is somewhat similar to the homodyne detection of squeezed state, there are
distinctions at two aspects: (1) the phenomenon is not affected by the quantum efficiency of the
detectors; (2) to have the maximum effect, the coherent source cannot be arbitrary -- it must have the
same two-photon rate as the two-photon source.

It should be pointed out that although some of the effects described in this paper have been observed in
a pulsed system \cite{koashi1}, some new features like the one shown in Fig.3(c) can only occur in a cw
system. Furthermore, pulse-pumped parametric down-conversion process is notorious for its temporal mode
mismatch which usually results in low visibility in interference. It relies on strong spectral filtering
\cite{lu} to achieve complete removal of the bunched photon pair to meet the requirement for the
application in quantum cryptography. For a cw system like ours, such a problem does not exist because we
can make time-resolved detection for narrow band fields.

In conclusion, we observed a variety of photon statistics ranging from photon bunching to
anti-bunching in two-photon interference between a coherent field and a
narrow-band two-photon field.  It is believed that we should be able to produce sub-Poissonian photon
statistics from this scheme. Such a system has the advantage over other atomic transition based
fluorescent system in that (1) it is not wavelength dependent so long as to meet phase matching
condition in parametric down-conversion; (2) it is more directional because of the phase matching
condition.

\acknowledgements
 This work was supported by the Office of Naval
Research.

\end{document}